\documentclass[aps,prl, amsmath, showpacs, preprintnumbers,superscriptaddress, twocolumn,sort&compress,floatfix, amssymb]{revtex4}
\usepackage{graphicx}
\usepackage{rotating}
\usepackage{dcolumn}
\usepackage{bm}
\usepackage{color}
\usepackage{mathptmx, textcomp}
\usepackage[latin1]{inputenc}
\usepackage{braket}

\begin{document}

\author{F. Meinert}
\affiliation{Institut f\"ur Experimentalphysik und Zentrum f\"ur Quantenphysik, Universit\"at Innsbruck, 6020 Innsbruck, Austria}
\author{M. J. Mark}
\affiliation{Institut f\"ur Experimentalphysik und Zentrum f\"ur Quantenphysik, Universit\"at Innsbruck, 6020 Innsbruck, Austria}
\author{E. Kirilov}
\affiliation{Institut f\"ur Experimentalphysik und Zentrum f\"ur Quantenphysik, Universit\"at Innsbruck, 6020 Innsbruck, Austria}
\author{K. Lauber}
\affiliation{Institut f\"ur Experimentalphysik und Zentrum f\"ur Quantenphysik, Universit\"at Innsbruck, 6020 Innsbruck, Austria}
\author{P. Weinmann}
\affiliation{Institut f\"ur Experimentalphysik und Zentrum f\"ur Quantenphysik, Universit\"at Innsbruck, 6020 Innsbruck, Austria}
\author{A. J. Daley}
\affiliation{Department of Physics and Astronomy, University of Pittsburgh, Pittsburgh, PA 15260, USA}
\author{H.-C. N\"agerl}
\affiliation{Institut f\"ur Experimentalphysik und Zentrum f\"ur Quantenphysik, Universit\"at Innsbruck, 6020 Innsbruck, Austria}

\title{Quantum quench in an atomic one-dimensional Ising chain}

\date{\today}

\pacs{37.10.Jk, 67.85.Hj, 75.10.Pq, 05.30.Rt}

\begin{abstract}
We study non-equilibrium dynamics for an ensemble of tilted one-dimensional atomic Bose-Hubbard chains after a sudden quench to the vicinity of the transition point of the Ising paramagnetic to anti-ferromagnetic quantum phase transition. The quench results in coherent oscillations for the orientation of effective Ising spins, detected via oscillations in the number of doubly-occupied lattice sites. We characterize the quench by varying the system parameters. We report significant modification of the tunneling rate induced by interactions and show clear evidence for collective effects in the oscillatory response.
\end{abstract}

\maketitle

Ultracold atomic ensembles confined in optical lattice potentials have proven to offer unique access to the study of strongly correlated quantum phases of matter \cite{Morsch2006,Bloch2008}. Unprecedented control over system parameters as well as exceptionally good isolation from the environment allow for implementation and quantitative simulation of lattice Hamiltonians \cite{Lewenstein2007,Sachdev2008}, not only bridging the fields of atomic and condensed matter physics in the study of ground-state phases, but also opening fundamentally new opportunities to explore out-of-equilibrium physics in essentially closed quantum systems \cite{Polkovnikov2011,Bloch2012}. For example, the rapid time-dependent control available over system parameters makes it possible to observe dynamics arising from a quantum quench, where a parameter such as the lattice depth is changed suddenly in time \cite{Cheneau2012, Trotzky2012, Will2010}.  Recently it was demonstrated that 1D chains of bosonic atoms with a superimposed linear gradient potential exhibit a quantum phase transition to a density-wave-ordered state, in which empty sites alternate with doubly-occupied sites (``doublons''). Beginning in a Mott-insulator phase of a Bose-Hubbard (BH) system \cite{Jaksch1998,Greiner2002}, where the on-site interactions dominate over tunneling and the atoms are exponentially localized on individual lattice sites, a gradient potential is added until the potential difference between adjacent sites matches the on-site interaction energy, and atoms can again resonantly tunnel. This was monitored for individual 1D chains with a length of about 10 sites with the quantum gas microscopy technique \cite{SIMON11}, and effectively maps onto a 1D Ising model \cite{SACHDEV02}, making it possible to simulate the transition from 1D paramagnetic (PM) spin chains to anti-ferromagnetic (AFM) spin chains in the context of ultracold atoms.

In this letter, we explore the dynamics of a quantum quench for bosonic atoms in such a tilted optical lattice \cite{RUBBO11,Kolodrubetz2012a,KOLOVSKY04}. Specifically, we quench the strength of the tilt to be near the phase transition point between PM and AFM regimes, and hence take the system far out of equilibrium, inducing strong oscillations in the number of doublons, which we detect through molecule formation. We find clear indications for the collective character of the ensuing dynamics.

\begin{figure}
\includegraphics[width=1\columnwidth]{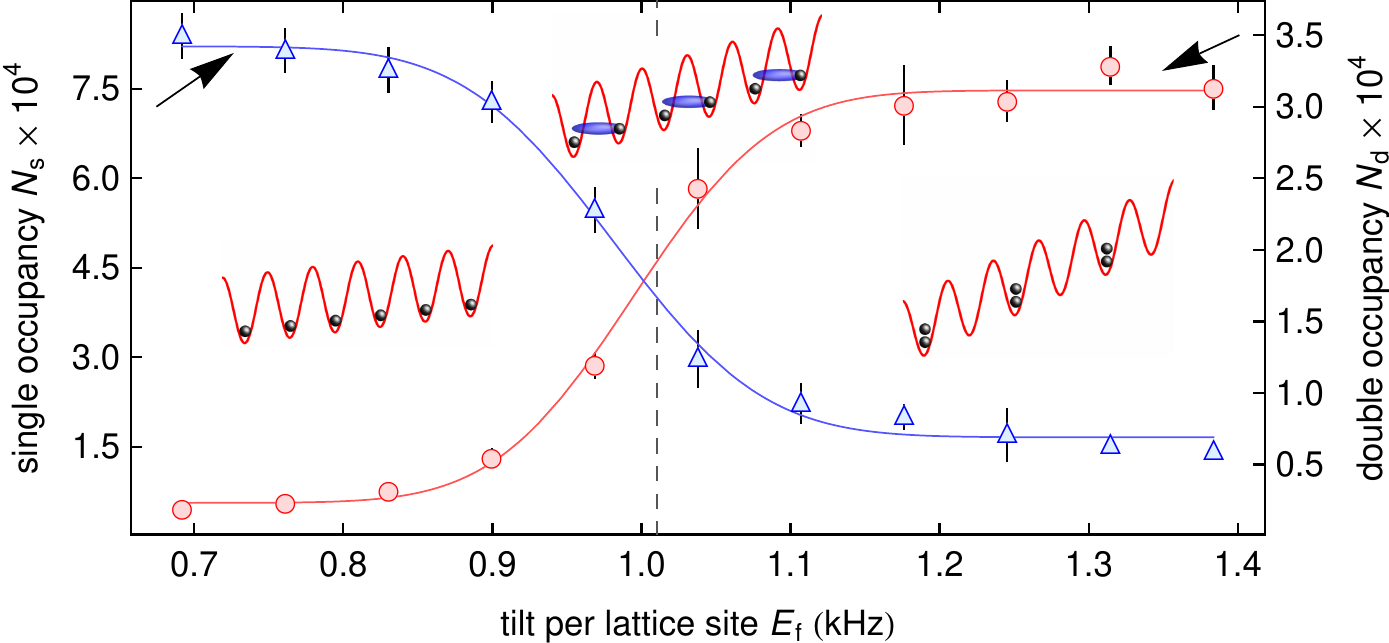}
\caption{\label{FIG1}(color online) Adiabatic tilt through the phase transition point. The number of doublons $N_\text{d}$ (circles) as detected through Feshbach molecule formation and the number of singly occupied sites $N_\text{s}$ (triangles) are plotted versus the tilt $E_\text{f}$ at the end of the adiabatic ramp for $V_z=10 \, \text{E}_\text{R}$ and $a_\text{s}=225(5)$ a$_0$, giving $U=0.963(20)$ kHz for $V_{y,z} =20 \, \text{E}_\text{R}$. The full ramp takes 40 ms. The solid lines are fits to guide the eye using the error function. The dashed line indicates $E_\text{c} =1.010(20)$ kHz. The insets show a schematic of the Ising phase transition for a 1D Mott-insulator chain from the PM (left) to the AFM (right) ground state \cite{SIMON11}. All error bars in this figure and in the following figures denote the one-sigma standard error.}
\end{figure}

We consider a 1D atomic ensemble in a tilted optical lattice potential near zero temperature. For sufficiently weak interaction energy, much smaller than the band gap, the system is described by the 1D single-band BH Hamiltonian \cite{Jaksch1998} augmented by a tilt \cite{SACHDEV02,KOLOVSKY04}
\begin{equation}
\hat{H} = -J \sum\limits_{\langle i,j \rangle } \hat{a}_i^\dagger \hat{a}_j + \sum\limits_{i} \frac{U}{2} \hat{n}_i\left(\hat{n}_{i}-1\right) +E \sum\limits_i i \hat{n}_{i}  + \sum\limits_{i} \epsilon_i \hat{n}_i \, .
\label{EQ1}
\end{equation}
As usual, $\hat{a}_i^\dagger$ ($\hat{a}_i$) are the bosonic creation (annihilation) operators at the $i$th lattice site, $\hat{n}_i = \hat{a}_i^\dagger \hat{a}_i$ are the number operators, $J$ is the tunnel matrix element, and $U$ is the on-site interaction energy. The linear energy shift from site to site is denoted by $E$, and $\epsilon_i$ accounts for a weak external confinement. For sufficiently small tilt ($E \ll U$) and sufficiently strong interactions ($U \gg J$) the ground state of the system for one-atom commensurate filling is a Mott insulator with exactly one atom per site. When the tilt is ramped from $E < U$ to $E > U$ across $E \approx U$ at finite $J$, the system establishes a regular periodic pattern of dipole states, for which one finds double occupancy at every second site with empty sites in between \cite{SACHDEV02,RUBBO11}, see insets to Fig.~\ref{FIG1}. The formation of such density-wave ordering is due to an effective nearest-neighbor constraint, which results in correlated tunneling of all particles in the 1D chain. When, in the course of the ramp, the tilt compensates for the interaction energy ($E \approx U$), the particles are allowed to resonantly tunnel onto the neighboring sites, however, only when the neighboring particle has not yet tunnelled itself. This system, when $|U-E|,J \ll U,E$, can be mapped onto an effective Ising spin model \cite{SACHDEV02,SIMON11}, where the two distinct ground states (PM and AFM order, respectively) are connected via a quantum phase transition with the quantum critical point $E_\text{c} = U + 1.85 J$ \cite{SACHDEVBOOK}.

\begin{figure}
\includegraphics[width=1\columnwidth]{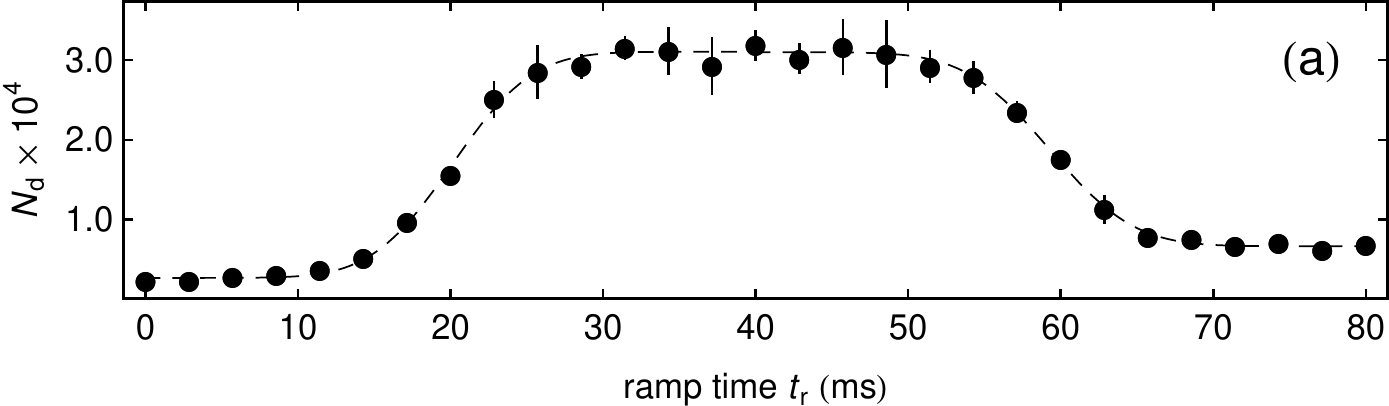}\\
\vspace{2mm}
\includegraphics[width=0.48\columnwidth]{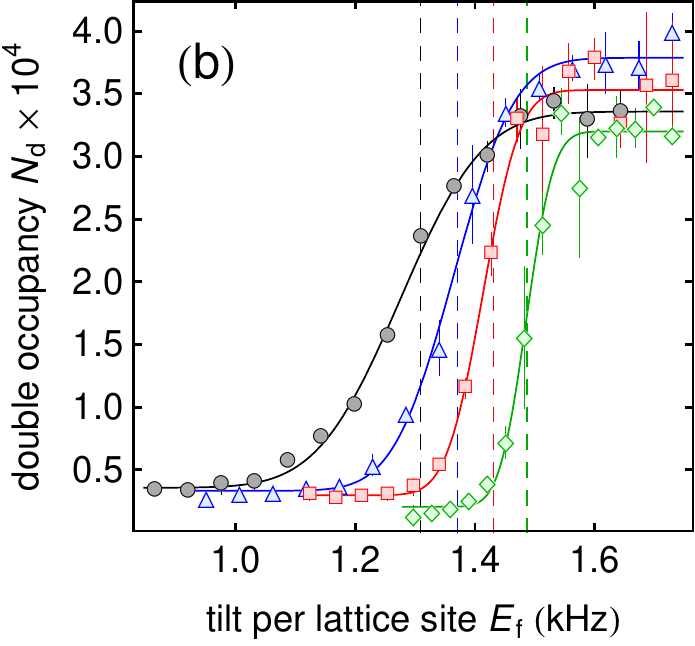}
\hspace{2mm}
\includegraphics[width=0.48\columnwidth]{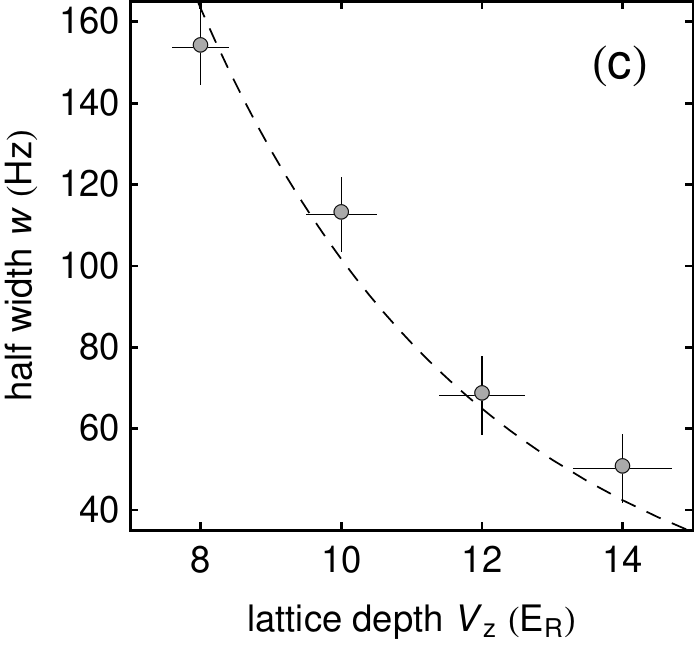}\\
\vspace{2mm}
\includegraphics[width=1\columnwidth]{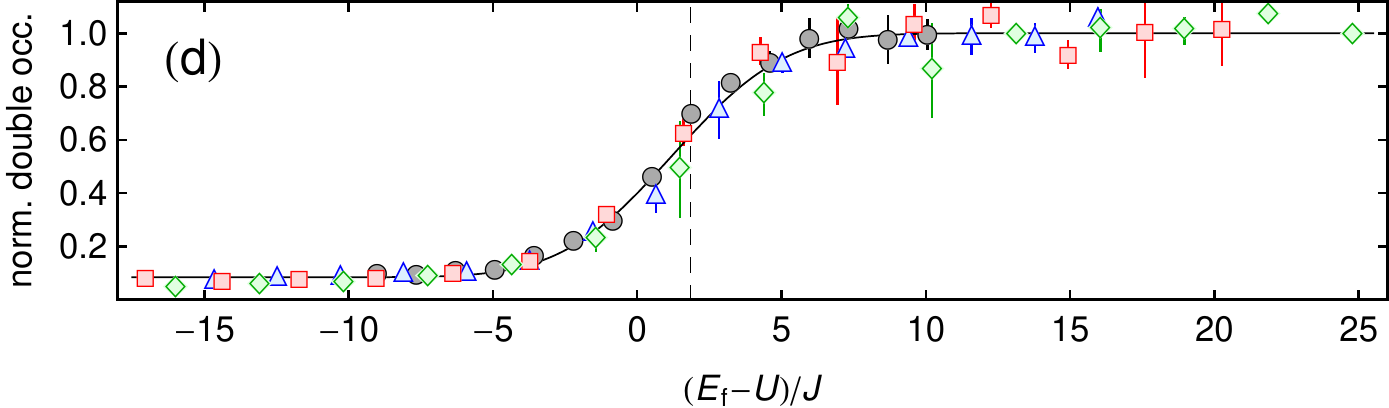}
\caption{\label{FIG2}(color online) (a) Reversibility of the phase transition. The number of doublons $N_\text{d}$ is plotted as the system is driven from the PM into the AFM ground state ($t \leq 40 \, \text{ms}$) and back ($t > 40 \, \text{ms}$) for $V_z = 10 \, \text{E}_\text{R}$. The dashed line is a fit using two concatenated error functions to guide the eye. For this measurement $V_{x,y} = 25 \, \text{E}_\text{R}$. (b) Phase transition for various values of the lattice depth $V_z = $ 8 (circles), 10 (triangles), 12 (squares), and $14 \, \text{E}_\text{R}$ (diamonds). The ramp rate is $\dot{E}=(17.3, 17.3, 8.6, 5.8)$ Hz/ms for $V_z=(8, 10, 12, 14) \, \text{E}_\text{R}$. The solid lines are fits using the error function to determine the half width $w$. For this measurement $V_{x,y} = 30  \, \text{E}_\text{R}$. We determine $U$ to $(1.233, 1.324, 1.401, 1.467)$ kHz and $E_\text{c}$ to $(1.309, 1.371, 1.431, 1.487)$ kHz for $V_z=(8, 10, 12, 14) \, \text{E}_\text{R}$ with typical error of $20$ Hz for $U$ and $E_\text{c}$. The positions of $E_\text{c}$ are indicated by the dashed lines. (c) Half width $w$ of the phase transition obtained from the fit as a function of $V_z$. The dashed line shows the calculated $4J$. (d) Same data as in (b), but normalized and plotted as a function of $ (E_\text{f}-U)/J $. The dashed line indicates the quantum critical point.}
\end{figure}

Starting with a Cs Bose-Einstein condensate \cite{Weber2002,Kraemer2004} we prepare an ensemble of 1D Bose chains by first creating a 3D Mott insulator in a cubic optical lattice with unity filling \cite{supmat}. Initially, the lattice depth $V_q$ is $20 \, \text{E}_\text{R}$ ($q=x,y,z$), where $\text{E}_\text{R}=1.325$ kHz \cite{units} is the photon recoil energy. The residual harmonic confinement along the vertical $z$-direction is $\nu_z = 11.9(0.2)~\rm{Hz}$. A tilt $E$ of up to $1.7$~kHz along $z$ is controlled by a magnetic force $|\nabla B|$. A Feshbach resonance allows us to control $U$ independent of $J$ \cite{Mark2011, Mark2012}.

We first drive the transition adiabatically from the PM to the AFM state. We do this in parallel for an ensemble of about 2000 1D systems ("tubes" or "Ising chains") with an average length of $40$ sites \cite{supmat}. We start with a pre-tilt of $E \approx 0.7 U$ and then lower $V_z$ to $10 \, \text{E}_\text{R}$ ($J=25.4$ Hz \cite{Jaksch1998}) within $10$ ms to allow for particle motion along the vertical direction. Effectively, our 3D one-atom-per-site Mott insulator has now turned into an ensemble of 1D tubes with near unity occupation as horizontal tunnel coupling between the tubes can be neglected on the timescale of the experiment. We ramp through the PM-to-AFM transition with a ramp speed of $\dot{E}=17.3$ Hz/ms up to a final value $E_\text{f}$ and project the ensemble of 1D many-body systems onto number states by quickly ramping $V_z$ up to its initial value and instantly reversing the tilt. We detect double occupancy through Feshbach molecule formation and detection \cite{Winkler2006,Danzl2010,supmat} with an overall doublon detection efficiency of $80(3)\%$. Fig.~\ref{FIG1} plots the number of doublons $N_\text{d}$ as a function of $E_\text{f}$. The transition from an ensemble of singly occupied sites to an ensemble with predominant double occupation can be seen clearly. The transition point is in very good agreement with the expectation from the BH model, with $U=0.963(20)$ kHz and $E_\text{c} =1.010(20)$ kHz. When ramping fully through the transition we create $N_\text{d}=3$ to $4 \times 10^4$ doublons, corresponding to ensembles with $79(2)\%$ double occupancy. We believe that defects in the initial one-atom Mott shell due to finite temperature and the superfluid outer ring are the main limitations to a higher conversion efficiency. We estimate the average length of AFM chains to $\approx 13$ lattice sites \cite{supmat}. Note that this state with exceptionally high double occupancy is ideal for creating ultracold samples of ground-state molecules \cite{Danzl2010}.

An unambiguous characteristic of a phase transition is its reversibility. Fig.~\ref{FIG2}(a) shows the result when the linear ramp (from 0 to 40 ms) is reversed (from 40 to 80 ms). We find that a large fraction of the doublons, more than $85\%$, is returned to the one-atom-per-site Mott phase.

\begin{figure}
\includegraphics[width=1\columnwidth]{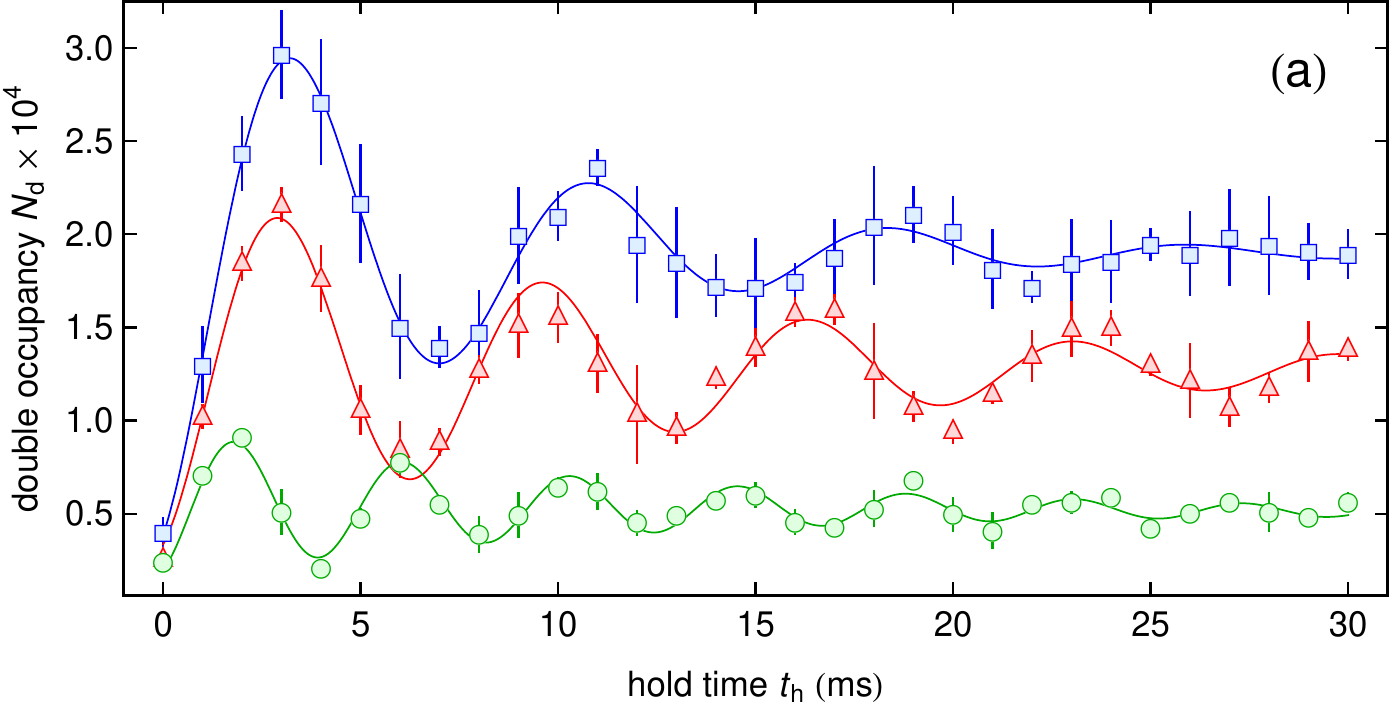}\\
\vspace{2mm}
\includegraphics[width=0.32\columnwidth]{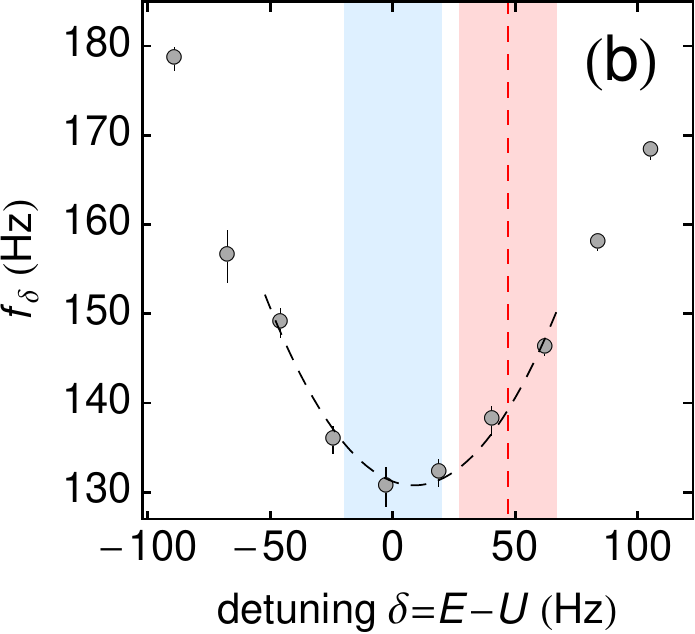}
\includegraphics[width=0.32\columnwidth]{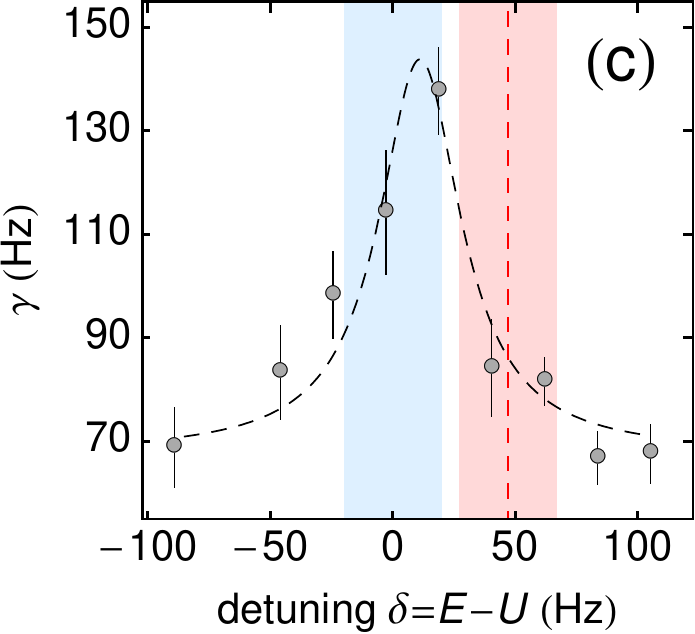}
\includegraphics[width=0.32\columnwidth]{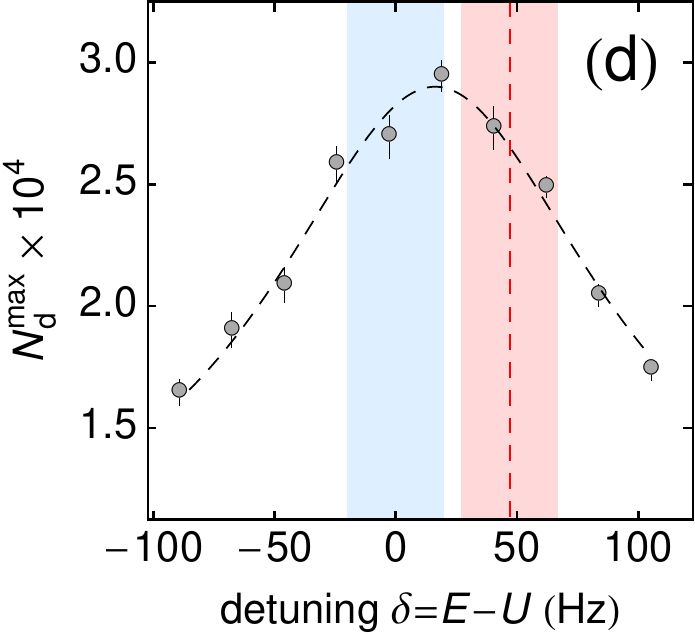}
\caption{\label{FIG3}(color online) Dynamic response for a quench to the resonance point $E \approx U$ with $U=1.019(20)$~kHz. (a) Number of doublons $N_\text{d}$ as a function of hold time $t_\text{h}$ for $E = 1.038$ (squares), $0.973$ (triangles), and $0.865$~kHz (circles). The solid lines are exponentially damped sinusoids fit to the data. (b) Extracted frequency $f_\delta$ as a function of $\delta=E-U$. The dashed line is a quadratic fit to the data in the vicinity of the minimum. (c) Extracted damping rate $\gamma$ as a function of $\delta$. The dashed line is a Lorentzian fit including a finite offset. (d) Extracted number of doublons $N_\text{d}^{\text{max}}$ at the first oscillation maximum as a function of $\delta$. The dashed line is a Lorentzian fit. The vertical dashed lines in (b)-(d) give the quantum critical point. The shaded areas (red and blue) indicate the one-sigma statistical experimental error (for $E_\text{c}-U$ and $\delta=0$, respectively). For all measurements $V_z = 10 \, \text{E}_\text{R}$, $V_{x,y} = 20 \, \text{E}_\text{R}$, and $a_\text{s}=238(5)$ a$_0$.}
\end{figure}

Adiabatic driving of the quantum phase transition from the PM into the AFM state is permitted by the finite tunneling rate $J$ between the sites through quantum fluctuations. In Fig.~\ref{FIG2}(b) we drive the phase transition at various values for the lattice depth $V_z$ in order to highlight the role of tunneling. To assure adiabaticity, we increase the ramp duration for larger $V_z$. Evidently, a reduced tunnel coupling for larger values of $V_z$ gives rise to a narrowing of the transition. Note that the transition point is shifted to a larger tilt as $V_z$ is increased due to the increase in $U$. We fit each data set with an error function of the form $\propto {\text{erf}} \left( (E-E_0) / w \right)$ and plot the extracted half-width $w$ as a function of $V_z$ in Fig.~\ref{FIG2}(c). We clearly find the width of the phase transition to scale with the tunnel matrix element. Interestingly, $w$ fits very well to the value $4 J$. In Fig.~\ref{FIG2}(d) we plot the data shown in Fig.~\ref{FIG2}(b) as a function of $ (E_\text{f}-U)/J $. The data collapses onto a single curve, confirming the role of $J$ as the natural energy scale \cite{SACHDEV02}.

Slow ramping through the critical point allows the system to follow adiabatically the many-body ground state. Non-equilibrium dynamics, however, arises when the initial Mott state is quenched by a rapidly applied tilt to the vicinity of the transition \cite{RUBBO11,KOLOVSKY04}. In our experiment we initiate non-equilibrium dynamics by first tilting the lattice and then quickly lowering $V_z$ to $10\text{E}_\text{R}$. We let the system evolve for a hold time $t_\text{h}$ and finally record the number of doublons $N_\text{d}$ as before. For the subsequent measurements $U=1.019(20)$ kHz. Fig.~\ref{FIG3}(a) shows the result for different values of $\delta=E-U$ close to the transition. The ensemble of 1D chains exhibits large amplitude oscillations for $N_\text{d}$. The amplitude is maximal for $\delta \approx 0$ and significantly lower for $|\delta|>0$. For $\delta \approx 0$, which we call the resonant case, the first maximum is reached at $t_\text{h} \approx 3$ ms. The oscillations damp out to an equilibrium value over a time scale of about 30 ms. We fit the data using exponentially damped sinusoids. From these we obtain the frequency $f_\delta$, the peak value of the first oscillation $N_\text{d}^{\text{max}}$, and the decay rate $\gamma$. Away from resonance, $N_\text{d}^{\text{max}}$ is lower, however $f_\delta$ is higher and, interestingly, $\gamma$ is lower. This can be seen from Fig.~\ref{FIG3}(b), (c), and (d), where we plot $f_\delta$, $\gamma$, and $N_\text{d}^{\text{max}}$ as a function of $\delta$. We fit $f_\delta$ by a quadratic function, as will be justified below, and obtain the position of the minimum $E_\text{res}$. The data for $N_\text{d}^{\text{max}}$ and $\gamma$ are fitted by Lorentzians. The damping is consistent with collapse of oscillations in coherent dynamics as observed in our simulations of the constrained Ising model \cite{supmat}, see also Refs.~\cite{SACHDEV02,KOLOVSKY04}. We find that the positions of the maxima for $N_\text{d}^{\text{max}}$ and $\gamma$ at $E=1035(2)$ and $E=1030(3)$ Hz, respectively, agree with the minimum for $f_\delta$ at $E_\text{res}=1028(1)$ Hz. Within our error bars \cite{errors}, all values are compatible with the value for $U$. We note that $N_\text{d}^{\text{max}}$ is only about $5\%$ lower than the value for $N_\text{d}$ we obtained from the adiabatic sweep as shown in Fig.~\ref{FIG1}.

The quadratic fit to the data for $f_\delta$ is justified, in the simplest approximation, by a two-state Rabi model for the case of a single double-well potential \cite{RUBBO11}. At the point $E \approx U$, the two Fock states $|11\rangle$ and  $|20\rangle$, describing one atom in each site and two atoms in the left site, respectively, are resonantly coupled by $J$, leading to Rabi oscillations between the two states with frequency $\sqrt{2} J$, where the factor $\sqrt{2}$ ensues from bosonic enhancement. The probability to find the system in the dipole state $|20\rangle$ thus oscillates in the case of finite detuning from resonance with the generalized Rabi frequency $f_\delta = \sqrt{8 J^2 + \delta^2}$ \cite{RUBBO11}, which is quadratic for sufficiently small $\delta$.

\begin{figure}
\includegraphics[width=0.48\columnwidth]{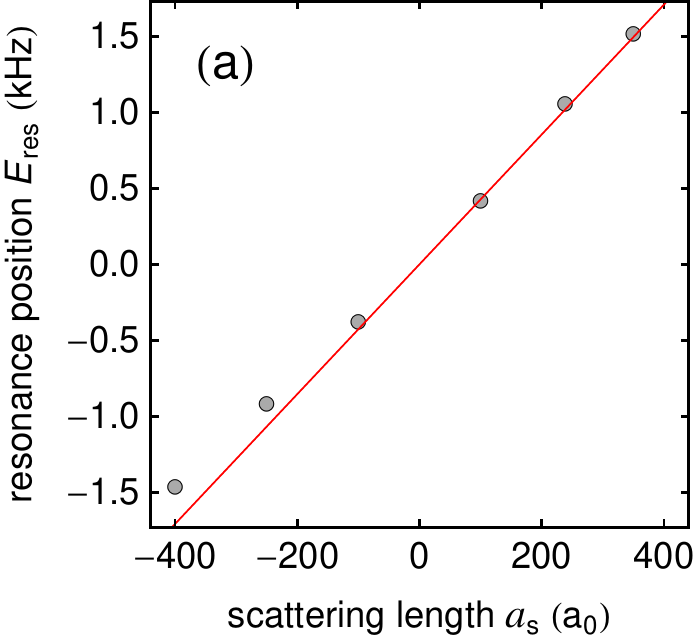}
\hspace{2mm}
\includegraphics[width=0.48\columnwidth]{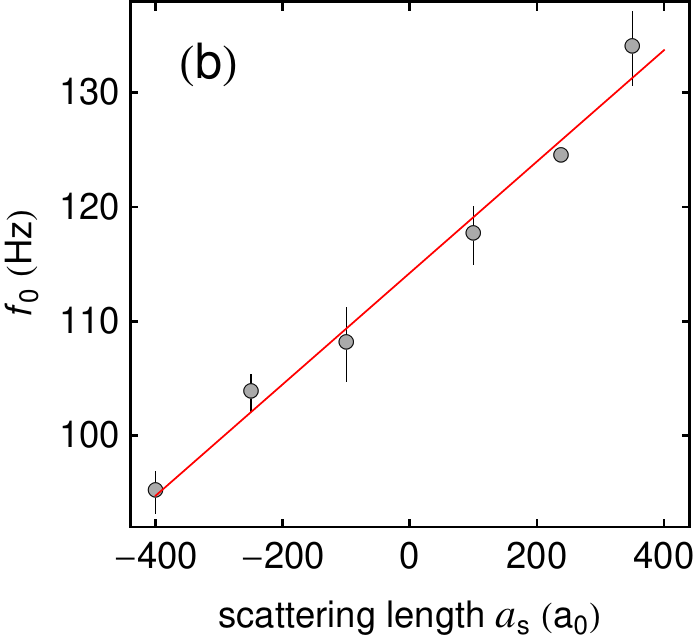}\\
\vspace{2mm}
\includegraphics[width=0.48\columnwidth]{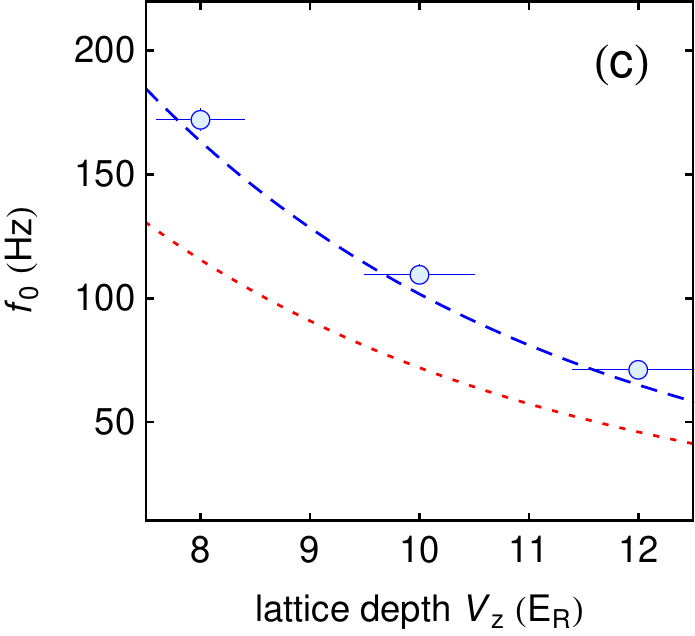}
\hspace{2mm}
\includegraphics[width=0.48\columnwidth]{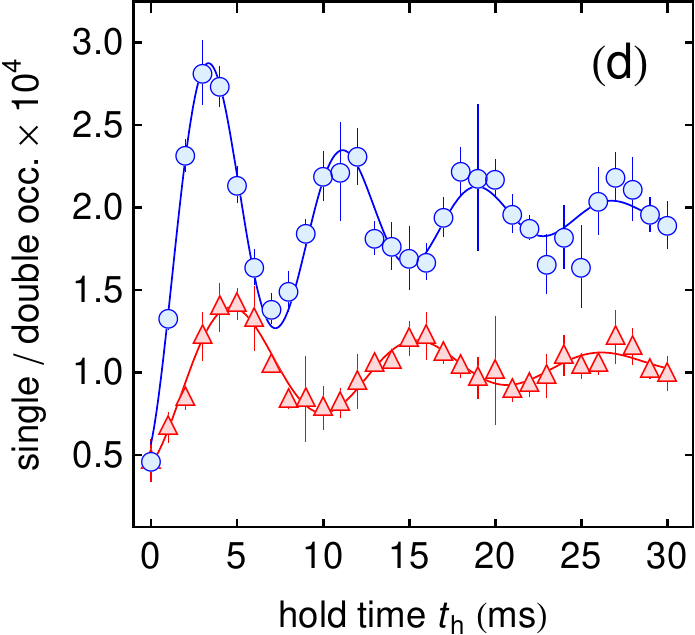}
\caption{\label{FIG4}(color online) Dependence of the resonance position $E_\text{res}$ (a) and of the on-resonance oscillation frequency $f_0$ (b) on $a_\text{s}$ as deduced from the dynamical response of the system. The solid line in (a) is the linear prediction of the BH model. In (b) the solid line is a linear fit to the data. For these measurements, $V_z= 10 \, \text{E}_\text{R}$ and $V_{x,y} = 20 \, \text{E}_\text{R}$. (c) On-resonance oscillation frequency $f_0$, interpolated to $a_\text{s}=0$, as a function of $V_z$ (circles). The dashed and dotted lines show the expected result for an Ising chain of three sites and the double well in the non-interacting limit, respectively. (d) On-resonance quench dynamics for the number of doublons in the 1D Ising chain (circles) and the number of singly occupied sites in nearly isolated double wells (triangles) for $U=1.060(20)$ kHz.}
\end{figure}

Before we continue this discussion we make use of our capability to tune $U$ by means of a Feshbach resonance. We repeat the measurements shown in Fig.~\ref{FIG3} for different values of $a_\text{s}$ to which we tune after producing the 3D one-atom Mott-insulator state and before initializing the quench. We determine the resonance position $E_\text{res}$ and the on-resonance oscillation frequency $f_0 = f_{\delta=0}$. The result is shown in Fig.~\ref{FIG4}(a) and (b). For repulsive interactions we find good agreement for $E_\text{res}$ with $U$ calculated from lowest-band Wannier functions \cite{Jaksch1998}. For attractive interactions there is a significant shift to lower absolute values of $E_\text{res}$ as corrections taking into account higher bands and the properly regularized pseudopotential would have to be included \cite{Mark2012}. We attribute the considerable dependence of the data for $f_0$ on $a_\text{s}$ to an effectively altered tunnel barrier as $U$ and hence $E_\text{res}$ are changed \cite{BISSBORT12,LUEHMANN12}.

We now quantitatively compare our findings for $f_0$ with the result from the simplified double-well system. Firstly, to remove the effects of the modified tunneling rate due to the tilt and interactions, we interpolate our data to find $f_0$ for $a_\text{s}=0$ with the linear fit as shown in Fig.~\ref{FIG4}(b). The resulting values are shown in Fig.~\ref{FIG4}(c) for data sets taken at three different values of $V_z$, and are compared to the prediction for the double well on resonance, $f_0=2\sqrt{2}J$ (dotted line). Evidently, the tunneling in the 1D chains is significantly faster when compared to the double-well system. This arises due to the nearest-neighbor constraint present in the 1D chains, which prohibits having two adjacent dipoles and which leads to an increase of the oscillation frequency even for three sites by a further factor of $\sqrt{2}$, to $4J$ (dashed line) \cite{RUBBO11}. Through numerical simulations with up to 60 lattice sites we have verified that the oscillation frequency shows only a slight further increase for longer chains beyond 3 sites \cite{supmat}. The weak harmonic trapping potential causes an additional small increase of the computed oscillation frequency. However, both effects are presently within our experimental uncertainty.

Finally, we compare the on-resonance oscillatory dynamics of the ensemble of 1D Ising chains with that of an ensemble of nearly isolated effective double-well systems to check the role of the constraint. The double wells are engineered by highly non-adiabatic loading of the 3D lattice, thereby creating a comparatively dilute sample of doublons. We clean out the singly occupied lattice sites by temporarily hiding the doublons in a molecular state and removing singular atoms by resonant light. We subsequently initiate the quench dynamics as before and detect the number of singly occupied sites \cite{supmat} as a function of $t_\text{h}$ as shown in Fig.~\ref{FIG4}(d). The oscillation frequencies in the two systems, determined to $f_0=128(2)$~Hz and $f_0=93(2)$~Hz, respectively, differ by a factor $1.38(3)$, in good agreement with the collective effect discussed above.

In conclusion, we have investigated quench dynamics by tuning a tilt suddenly onto the transition from PM to AFM Ising order in a large ensemble of 1D atomic Mott-insulator chains. We observe significant shifts of the tunneling parameter arising from interactions, and collective effects in the oscillations arising from the effective constraint in the model. This study of quench dynamics opens the opportunity to explore many aspects of the dynamics in these systems that up to now were only addressed theoretically. This includes scaling relations for smaller quenches across the phase transition \cite{Kolodrubetz2012a,Kolodrubetz2012b,DeGrandi2010}, as well as possibly quenches at other resonance points (e.g., near $E=U/2$).

We are indebted to R. Grimm for generous support, and thank J. Schachenmayer for discussions and contributions to numerical code development. We gratefully acknowledge funding by the European Research Council (ERC) under Project No. 278417, and support in Pittsburgh from NSF Grant PHY-1148957.

\bibliographystyle{apsrev}

\newpage
\clearpage

\section{Supplementary Material: Quantum quench in an atomic one-dimensional Ising chain}

\subsection{Preparation of the one-atom-per-site Mott insulator}

To prepare the ensemble of 1D Bose chains we start with a 3D Bose-Einstein condensate (BEC) without detectable non-condensed fraction of typically $8.5 \times 10^4$ Cs atoms in the energetically lowest hyperfine ground state $|F\!=\!3,m_F\!=\!3\rangle$ confined to a crossed dipole trap. Trapping and cooling procedures are described in Refs.~\cite{Weber2002b,Kraemer2004b}. The sample is levitated against gravity by a vertical magnetic field gradient of $|\nabla B| \approx 31.1 \, \rm{G/cm}$. The BEC is adiabatically loaded from the trap into a cubic 3D optical lattice generated by three mutually perpendicular retro-reflected laser beams at a wavelength of $\lambda_\text{l}=1064.5$~nm to induce the phase transition to a 3D Mott insulator. The final value for the lattice depth $V_q$ in each direction ($q=x,y,z$) is typically $20 \, \text{E}_\text{R}$. For some of the experiments the depth is set to a higher value. Here, $\text{E}_\text{R}=1.325$ kHz denotes the photon recoil energy for Cs atoms at $\lambda_\text{l}$. We take care to prepare the system with a clean singly occupied Mott shell with less than $4\%$ of residual double occupancy. When the dipole trap is extinguished the residual harmonic confinement in the direction of gravity as a result of the transversal profile of the lattice beams is $\nu_z = 11.9(0.2)~\rm{Hz}$. A tilt up to $E=1.7$~kHz along the vertical $z$-direction can be introduced by lowering the strength of the levitation field. A broad Feshbach resonance allows us to precisely tune the atomic scattering length $a_\text{s}$ and thus to set $U$ \cite{Mark2011b, Mark2012b}. The day-to-day variations in the number of particles in the BEC lead to variations in the number of doublons $N_\text{d}$, as can be seen in e.g. Fig. 2(b). There, the different data sets have been recorded on different days.

\subsection{Calibration of $E$ and $U$}

\begin{figure}
\includegraphics[width=1\columnwidth]{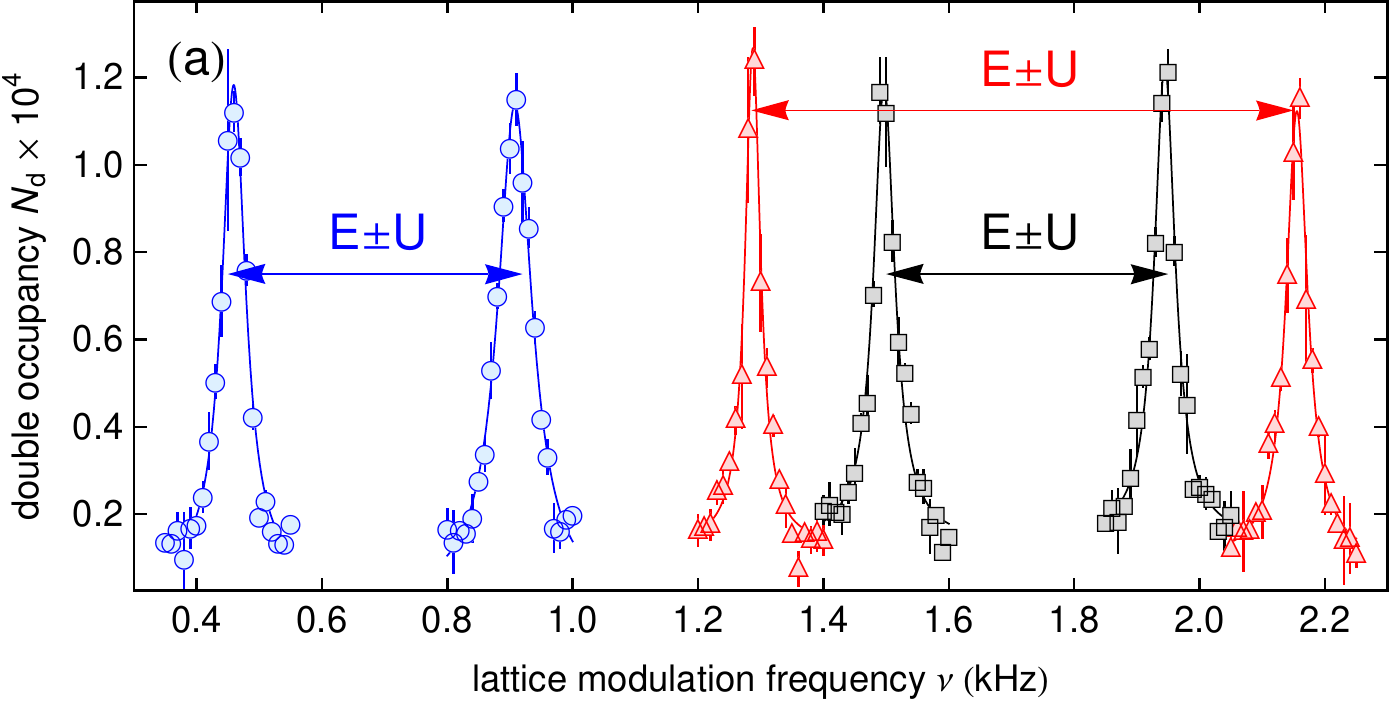}\\
\vspace{2mm}
\includegraphics[width=0.48\columnwidth]{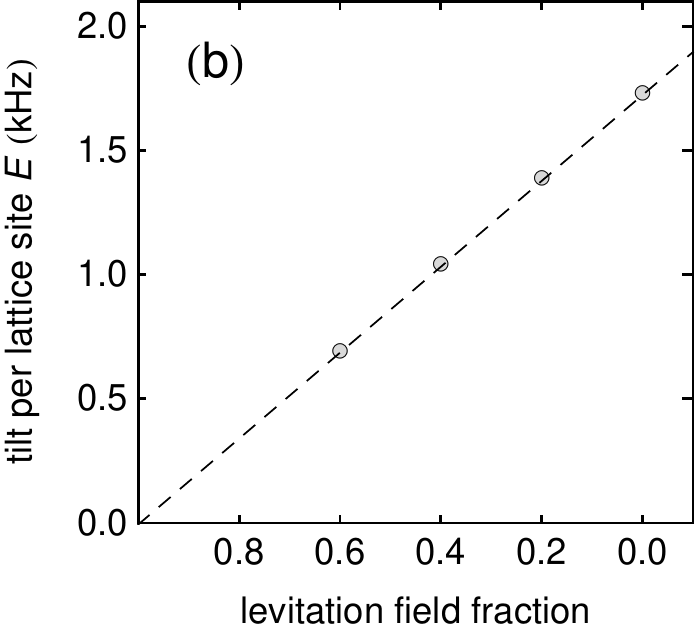}
\hspace{2mm}
\includegraphics[width=0.48\columnwidth]{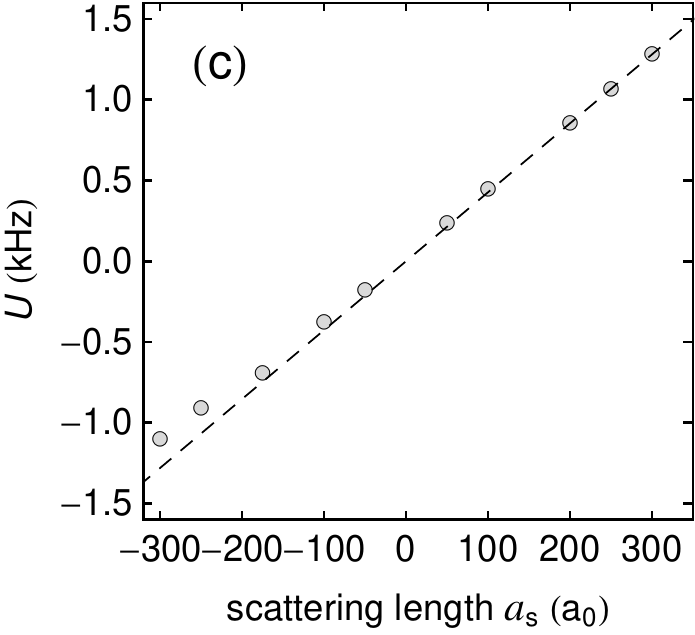}
\caption{\label{FIGAPP1}(color online)  (a) Typical lattice modulation spectra showing tunneling resonances at modulation frequencies $E \pm U$ as detected by an increase in the number of doublons $N_\text{d}$. We have set $a_\text{s}=50$~a$_0$ and $E=1.73$~kHz (squares), $a_\text{s}=100$~a$_0$ and $E=1.73$~kHz (triangles), and $a_\text{s}=50$~a$_0$ and $E=0.69$~kHz (circles). The solid lines are Lorentzian fits to determine the center position of the resonances. (b) Tilt per lattice site $E$ as a function of the fraction of the magnetic levitation field. The dashed line is a linear fit to the data. The data is taken at $a_\text{s}=50$~a$_0$. (c) On-site interaction energy $U$ as a function of $a_\text{s}$. The prediction from the BH model is depicted by the dashed line. In (b) and (c) all error bars are smaller than the data points.}
\end{figure}

The tilt $E$ is set by reducing the vertical magnetic field gradient to a certain fraction. We calibrate $E$ by lattice modulation spectroscopy. For the calibration measurements we set $E$ to a value far detuned from the resonance point $E_\text{res}$ such that tunneling remains suppressed by the on-site interaction. Modulation of $V_z$ with a frequency $\nu_\pm = E \pm U$ bridges the energy gap and results in driven tunneling along and against the applied potential gradient onto neighboring sites \cite{Ma2011}.

We modulate the lattice sinusoidally for $100$ ms with an amplitude of typically $5\%$ around its mean value $V_z = 10\text{E}_\text{R}$. Three typical excitation spectra in the number of doublons created are shown in Fig.~\ref{FIGAPP1}(a). In each spectrum we identify two strong and narrow tunneling resonances at $\nu_\pm = E \pm U$. The resonance positions provide $E$ and $U$. Keeping $a_\text{s}$ constant they allow for precise calibration of $E$ as a function of the magnetic levitation field, as shown in Fig.~\ref{FIGAPP1}(b). Analogous calibration of $U$ as a function of $a_\text{s}$ for a fixed tilt is depicted in Fig.~\ref{FIGAPP1}(c). The data are in very good agreement with the prediction from the BH model \cite{Jaksch1998b} for $a_\text{s} > 0$, whereas for $a_\text{s} < 0$, $U$ is considerably shifted to smaller absolute values due to multi-body interaction effects \cite{Mark2012b}.

\subsection{Estimating the average chain length}

\begin{figure}
\includegraphics[width=1\columnwidth]{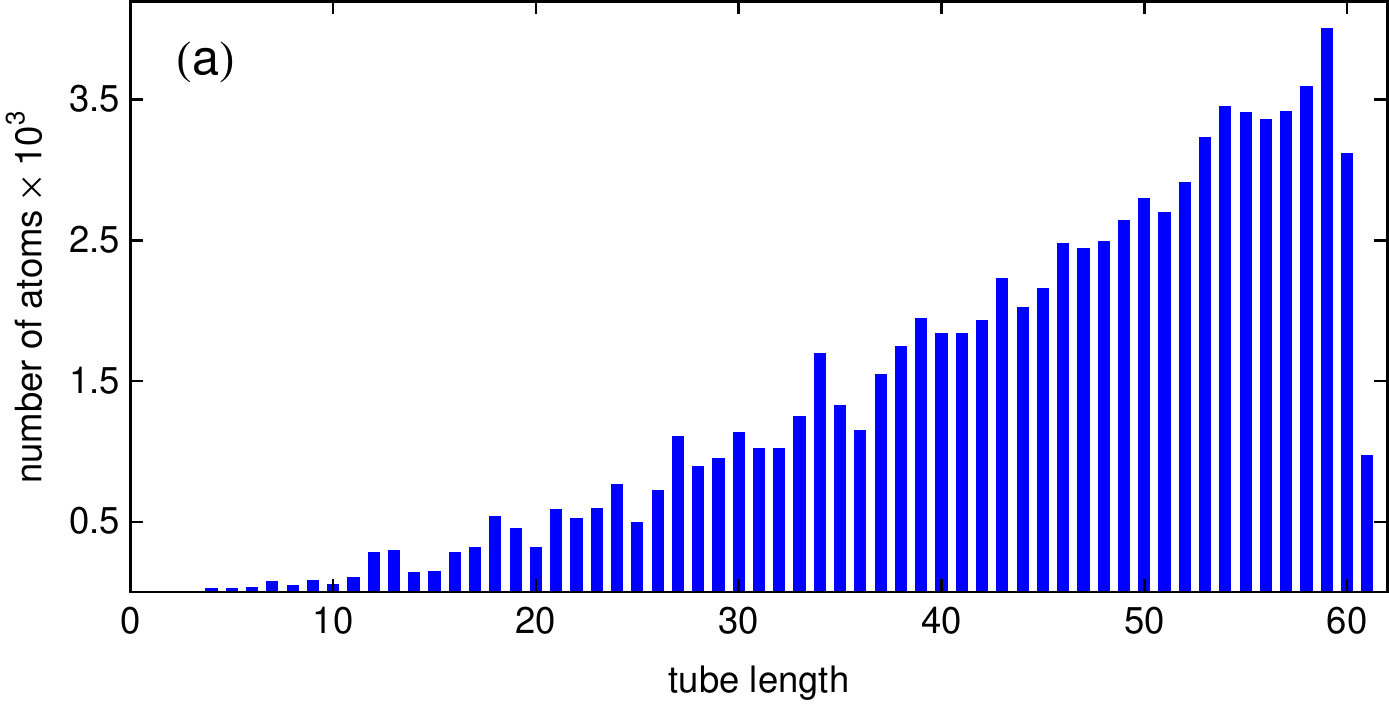}\\
\vspace{2mm}
\includegraphics[width=1\columnwidth]{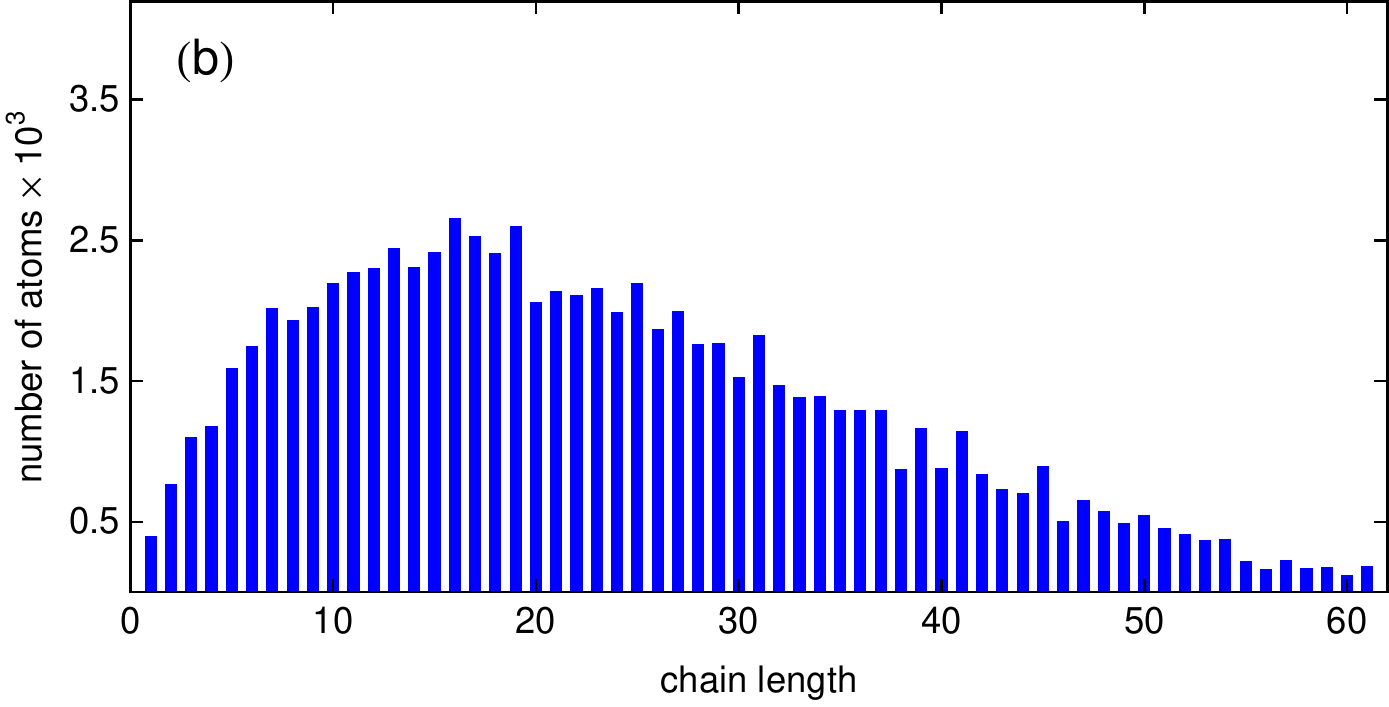}
\caption{\label{FIGAPP2}(color online) (a) Calculated number of atoms as a function of tube length using the Gutzwiller mean field method. (b) Calculated number of atoms as a function of chain length after introducing $5\,\%$ randomly distributed defects. The resulting average chain length is calculated to $\approx13$ lattice sites.}
\end{figure}

To get a rough estimate for the average chain length for our ensemble of tubes we first numerically determine the ground state of the 3D Mott insulator using the Gutzwiller mean field method \cite{Gutzwiller1963}, feeding in our typical values for the particle number, the trapping strength, and the interaction strength. Fig.~\ref{FIGAPP2}(a) shows the calculated atom number distribution. The average atom number per tube is $\approx40$. Defects are introduced by removing atoms from lattice sites at random positions. The remaining uninterrupted chains with a certain length can then be counted. Fig.~\ref{FIGAPP2}(b) shows the resulting distribution when introducing $5\,\%$ defects. The average chain length decreases to $\approx13$ lattice sites, but evidently a reasonable number of longer chains survives.

\subsection{Detection of single and double occupancy}

We detect double occupancy by associating atom pairs on a lattice site to weakly bound molecules crossing a narrow g-wave Feshbach resonance with a pole at $19.8$~G \cite{MARK07}. The system is cleaned from remaining singly occupied sites by combining rapid adiabatic passage from $|F\!=\!3,m_F\!=\!3\rangle$ to $|F\!=\!4,m_F\!=\!4\rangle$ using a microwave field with a resonant light pulse \cite{Danzl2010b}. After dissociating the molecules again we detect the number of atoms with standard absorption imaging. Single occupancy is measured by associating doubly occupied sites into Feshbach molecules, thereby hiding these during the imaging process.

Our measurement noise is mainly affected by atom number fluctuations in the BEC. When detecting double occupancy, additional noise arises from small shot-to-shot variations in the molecule association, cleaning, and dissociation efficiency. Especially for experiments with comparatively low absolute atom numbers, detection of single occupancy is favored as noise from the cleaning and dissociation process is absent.

\subsection{Numerical simulations of quench dynamics}

\begin{figure}
\includegraphics[width=1\columnwidth]{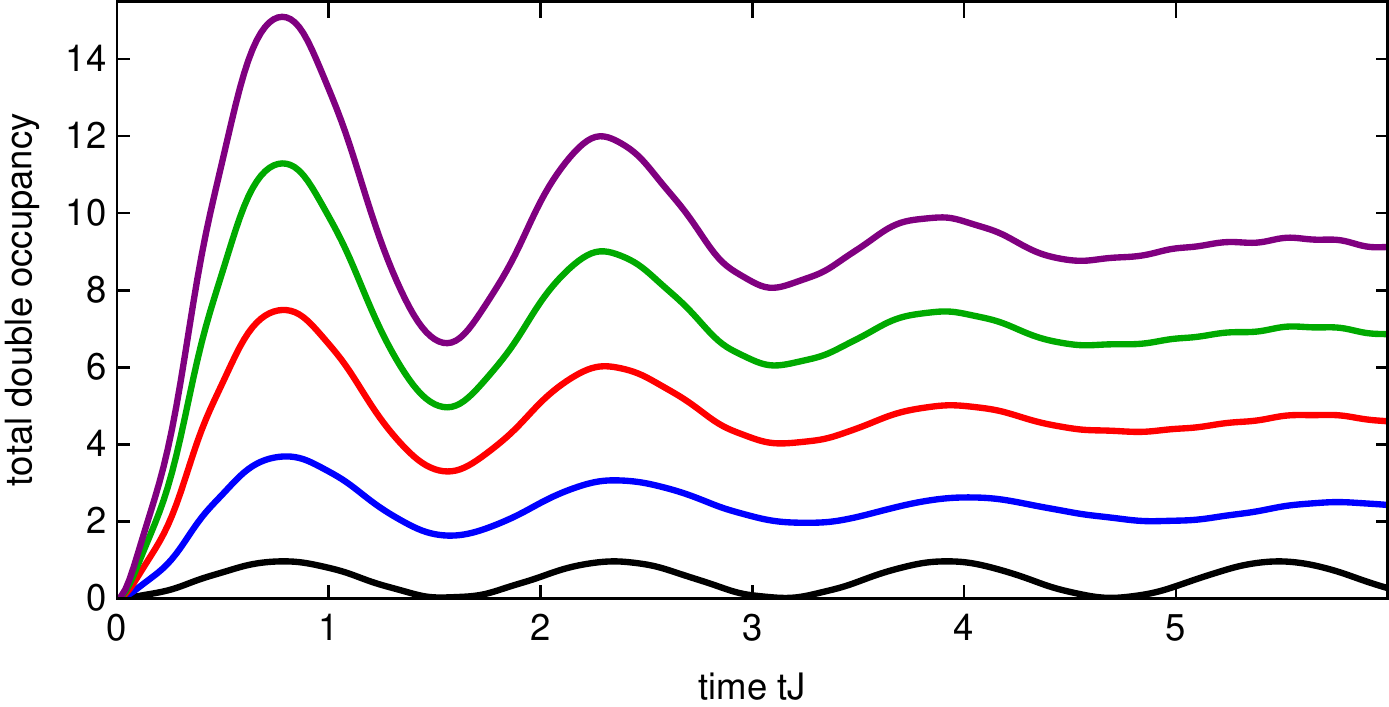}
\caption{\label{FIGAPP3}(color online) Number of doubly occupied sites as a function of time $tJ$ ($\hbar=1$), as computed from the Bose-Hubbard model beginning from an initial state with one particle per lattice site. Here we choose $E=U=12J$, and vary the number of lattice sites $N=3$ (black), 10 (blue), 20 (red), 30 (green), and 40 (purple). These calculations are performed with t-DMRG methods, which are converged for matrix sizes $D=100$--$200$.} 
\end{figure}

We performed numerical simulations of the quench dynamics both within the effective contrained Ising model, and directly using a single-band Bose-Hubbard model. Using combinations of exact diagonalization techniques and time-dependent density matrix renormalization group (t-DMRG) methods \cite{Vidal04,tdmrg1,tdmrg2,dmrgrev}, we calculated the propagation in time, beginning with an initial state with one particle per lattice site. Example calculations for the number of doubly-occupied sites as a function of time with different numbers of sites in a Bose-Hubbard model with open boundary conditions are shown in Fig.~\ref{FIGAPP3}.

There are several features that we recognize from these calculations. First, we note that although collective effects for chains longer than $N=3$ lead to higher frequency components appearing in the oscillations, these do not substantially affect the period of oscillations over the first few cycles - as used in the manuscript to determine the frequencies in Fig.~4 (note that in the experiment, we implicitly average these values over chains of different length). These high-frequency components that enter due to collective effects instead lead to dephasing of the oscillations, as can be seen in the curves for systems larger than three sites. For large chain lengths $N\gtrsim 30$, this dephasing, combined with some averaging over chains of different length, is sufficient to account for the rate of damping seen in the experiment. In practice, other sources of dephasing, e.g., due to the harmonic confinement or noise, will also contribute to damping in the experiment. However, we estimate in our experiments that the dominant contribution arises from these coherent collective effects.

\bibliographystyle{apsrev}

\begin{references}

\bibitem{Morsch2006} O. Morsch and M. Oberthaler, Rev. Mod. Phys. \textbf{78},179 (2006).

\bibitem{Bloch2008} I. Bloch, J. Dalibard, and W. Zwerger, Rev. Mod. Phys. \textbf{80}, 885 (2008).

\bibitem{Lewenstein2007} M. Lewenstein, A. Sanpera, V. Ahufinger, B. Damski, A. Sen De, and U. Sen, Adv. Phys. \textbf{56}, 243 (2007).

\bibitem{Sachdev2008} S. Sachdev, Nature Phys. \textbf{4}, 173 (2008).

\bibitem{Polkovnikov2011} A. Polkovnikov, K. Sengupta, A. Silva, and M. Vengalattore, Rev. Mod. Phys. {\bf 83}, 863 (2011).

\bibitem{Bloch2012} I. Bloch, J. Dalibard and S. Nascimbne, Nat. Phys. {\bf 8}, 267 (2012).

\bibitem{Cheneau2012} M. Cheneau, P. Barmettler, D. Poletti, M. Endres, P. Schau\ss, T. Fukuhara, C. Gross, I. Bloch, C. Kollath, and S. Kuhr, Nature {\bf 481},  484 (2012).

\bibitem{Trotzky2012} S. Trotzky, Y-A. Chen, A. Flesch, I. P. McCulloch, U. Schollw\"ock, J. Eisert, and I. Bloch, Nature Phys. {\bf 8}, 325 (2012).

\bibitem{Will2010} S. Will, T. Best, U. Schneider, L. Hackerm\"uller, D- S. L\"uhmann, and I. Bloch, Nature {\bf 465}, 197 (2010).

\bibitem{Jaksch1998} D. Jaksch, C. Bruder, J.I. Cirac, C.W. Gardiner, and P. Zoller, Phys. Rev. Lett. \textbf{81}, 3108 (1998).

\bibitem{Greiner2002} M. Greiner, O. Mandel, T. Esslinger, T. W. H\"{a}nsch, and I. Bloch, Nature \textbf{415}, 39 (2002).

\bibitem{SIMON11} J. Simon, W. S. Bakr, R. Ma, M. E. Tai, P. M. Preiss, and M. Greiner, Nature \textbf{472}, 307 (2011).

\bibitem{SACHDEV02} S. Sachdev, K. Sengupta, and S. M. Girvin, Phys. Rev. B \textbf{66}, 075128 (2002).

\bibitem{RUBBO11} C. P. Rubbo, S. R. Manmana, B. M. Peden, M. J. Holland, and A. M. Rey, Phys. Rev. A {\bf{84}}, 033638 (2011).

\bibitem{Kolodrubetz2012a} M. Kolodrubetz, D. Pekker, B. K. Clark, and K. Sengupta, Phys. Rev. B {\bf 85}, 100505 (2012).

\bibitem{KOLOVSKY04} A. R. Kolovsky, Phys. Rev. A {\bf{70}}, 015604 (2004).

\bibitem{SACHDEVBOOK} S. Sachdev, \textit{Quantum Phase Transitions } (Cambridge University Press, Cambridge, England, 1999).

\bibitem{Weber2002} T. Weber, J. Herbig, M. Mark, H.-C. N\"agerl, and R. Grimm, Science \textbf{299}, 232 (2002).

\bibitem{Kraemer2004} T. Kraemer, J. Herbig, M. Mark, T. Weber, C. Chin, H.-C. N\"agerl, and R. Grimm, Appl. Phys. B \textbf{79}, 1013 (2004).

\bibitem{supmat} See Supplemental Material at [URL will be inserted by publisher]

\bibitem{units} We give all energies in frequency units.

\bibitem{Mark2011} M. J. Mark, E. Haller, K. Lauber, J. G. Danzl, A. J. Daley, H.-C. N\"agerl, Phys. Rev. Lett. \textbf{107}, 175301 (2011).

\bibitem{Mark2012}M. J. Mark, E. Haller, K. Lauber, J. G. Danzl, A. Janisch, H. P. B\"uchler, A. J. Daley, and H.-C. N\"agerl, Phys. Rev. Lett. \textbf{108}, 215302 (2012).

\bibitem{Winkler2006} K. Winkler, G. Thalhammer, F. Lang, R. Grimm, J. Hecker Denschlag, A. J. Daley, A. Kantian, H. P. B\"uchler, and P. Zoller, Nature \textbf{441}, 853 (2006).

\bibitem{Danzl2010} J. G. Danzl, M. J. Mark, E. Haller, M. Gustavsson, R. Hart, J. Aldegunde, J. M. Hutson, and H.-C. N\"agerl, Nature Physics \textbf{6}, 265 (2010).

\bibitem{errors} We note that the day-to-day variation of the positions of the extrema for $N_\text{d}^{\text{max}}$, $\gamma$, and $f_\delta$ is about 10 Hz and hence considerably larger than the statistical error from the fits.

\bibitem{BISSBORT12}U. Bissbort, F. Deuretzbacher, and W. Hofstetter, Phys. Rev. A \textbf{86}, 023617 (2012).

\bibitem{LUEHMANN12}D. S. L\"uhmann, O. J\"urgensen, and K. Sengstock, New J. Phys. \textbf{14}, 033021 (2012).

\bibitem{Kolodrubetz2012b} M. Kolodrubetz, B. K. Clark, and D. A. Huse, Phys. Rev. Lett. {\bf 109}, 015701 (2012).

\bibitem{DeGrandi2010} C. De Grandi, V. Gritsev, and A. Polkovnikov, Phys. Rev. B {\bf 81}, 012303 (2010).

\end{references}

\begin{references}
\bibitem{Weber2002b} T. Weber, J. Herbig, M. Mark, H.-C. N\"agerl, and R. Grimm, Science \textbf{299}, 232 (2002).
\bibitem{Kraemer2004b} T. Kraemer, J. Herbig, M. Mark, T. Weber, C. Chin, H.-C. N\"agerl, and R. Grimm, Appl. Phys. B \textbf{79}, 1013 (2004).
\bibitem{Mark2011b} M. J. Mark, E. Haller, K. Lauber, J. G. Danzl, A. J. Daley, H.-C. N\"agerl, Phys. Rev. Lett. \textbf{107}, 175301 (2011).
\bibitem{Mark2012b}M. J. Mark, E. Haller, K. Lauber, J. G. Danzl, A. Janisch, H. P. B\"uchler, A. J. Daley, and H.-C. N\"agerl, Phys. Rev. Lett. \textbf{108}, 215302 (2012).
\bibitem{Ma2011}R. Ma, M. E. Tai, P. M. Preiss, W. S. Bakr, J. Simon, and M. Greiner, Phys. Rev. Lett. {\bf{107}}, 095301 (2011).
\bibitem{Jaksch1998b} D. Jaksch, C. Bruder, J.I. Cirac, C.W. Gardiner, and P. Zoller, Phys. Rev. Lett. \textbf{81}, 3108 (1998).
\bibitem{Gutzwiller1963}M. C. Gutzwiller, Phys. Rev. Lett. {\bf{10}}, 159 (1963).
\bibitem{MARK07}M. Mark, F. Ferlaino, S. Knoop, J. G. Danzl, T. Kraemer, C. Chin, H.-C. N\"agerl, and R. Grimm, Phys. Rev. A {\bf{76}}, 042514 (2007).
\bibitem{Danzl2010b} J. G. Danzl, M. J. Mark, E. Haller, M. Gustavsson, R. Hart, J. Aldegunde, J. M. Hutson, and H.-C. N\"agerl, Nature Physics \textbf{6}, 265 (2010).
\bibitem{Vidal04} G. Vidal, Phys. Rev. Lett. {\bf{93}}, 040502 (2004).
\bibitem{tdmrg1} A. J. Daley, C. Kollath, U. Schollw\"ock, G. Vidal, J. Stat. Mech.: Theor. Exp. P04005 (2004).
\bibitem{tdmrg2} S. R. White and A. E. Feiguin, Phys. Rev. Lett. {\bf{93}}, 076401 (2004).
\bibitem{dmrgrev} F. Verstraete, V. Murg, and J. I. Cirac, Adv. Phys. {\bf{57}}, 143 (2008).

\end{references}

\clearpage

\end{document}